\pdfoutput=1
\RequirePackage{fix-cm}
\documentclass{svjour3}                     
\smartqed  
 \usepackage{mathptmx}      
\usepackage{graphicx}   
\usepackage{float}
\usepackage{wrapfig}
\usepackage{amsmath,amssymb}
\begin{document}

\title{
A Realistic Approach to the $\Xi NN$ Bound-State Problem based on Faddeev Equation
}


\author{K. Miyagawa         \and
              M. Kohno 
}


\institute{K. Miyagawa \at
              Graduate School of Science, Okayama University of Science, Okayama 700-0005, Japan \\
             \email{miyagawa@rcnp.osaka-u.ac.jp}    \\
             \emph{Present address:} Research Center for Nuclear Physics, Osaka University, Ibaraki 567-0047, Japan  
           \and
           M. Kohno \at
              Research Center for Nuclear Physics, Osaka University, Ibaraki 567-0047, Japan \\
              \email{kohno@rcnp.osaka-u.ac.jp}    \\
}

\date{Received: date / Accepted: date}

\maketitle

\begin{abstract}
The Faddeev equations for the $\Xi NN$ bound-state problem are solved where  the three $S$=$-2$ baryon-baryon interactions of
J\"ulich-Bonn-M\"unchen chiral EFT, HAL QCD and Nijmegen ESC08c are used.
The $T$-matrix $T_{\Xi N, \Xi N}$ obtained within the original 
$\Lambda\Lambda$-$\Xi N$-$\Sigma\Sigma$\,$/$\,$\Xi N$-$\Lambda \Sigma$-$\Sigma\Sigma$
coupled-channel framework is employed as an input to the equations. We found no bound state for
  J\"ulich-Bonn-M\"unchen chiral EFT and  HAL QCD
but ESC08c generates a bound state with the total isospin and  spin-parity
$(T,J^{\pi})=(1/2, 3/2^+)$ where the decays into  $\Lambda\Lambda N$ are suppressed. 
\end{abstract}

\section{Introduction}
\label{intro}
   In the last decade, the description of  $S$=$-2$ baryon-baryon interactions  has been significantly developed;  
in addition to conventional meson-theoretical approaches,  the ways established on chiral effective theory and lattice 
QCD simulation have made remarkable progress.
  Following this development,  analyses of three-baryon systems with $S$=$-2$ have appeared~\cite{Gar,Fil,Hiy}.
The interactions adopted there are, however,  more or less simplified, and  the results obtained appear to be still  primitive.
  This paper  presents an analysis  of the $\Xi NN$ system as a bound state on the basis of  Faddeev equation, which
 uses the three descriptions of the interaction  for the $\Lambda\Lambda$-$\Xi N$-$\Sigma\Sigma$
 and $\Xi N$-$\Lambda \Sigma$-$\Sigma\Sigma$ coupled systems: 
J\"ulich-Bonn-M\"unchen chiral EFT (J\"ulich Ch-EFT)~\cite{JEF,JEF2,JEF3},  
 HAL QCD~\cite{HAL,HAL2}, and Nijmegen ESC08c~\cite{ES8}.     

An advantage  of the Faddeev approach is  that the inputs to the equations are two-body $T$-matrices. 
Notice that the $\Lambda\Lambda$, $\Xi N$ and $\Sigma\Sigma$ systems are coupled in  $^1S_0$ for $t \,$(isospin)=$0$
states, while $\Xi N$  is not coupled to other channels  in $^3S_1$-$^3D_1$.
On the other hand,  for  $t=$1,   $\Xi N$ and $\Lambda \Sigma$ are coupled in $^1S_0$, while $\Xi N$,  $\Lambda \Sigma$
and $\Sigma\Sigma$ are coupled in $^3S_1$-$^3D_1$.
 After precisely solving  these two-body coupled-channel problems,  we use the $T$-matrix $T_{\Xi N, \Xi N}$ as the input
to $\Xi NN$ Faddeev equations.
Although  entire couplings in the three-body space are not included,  this usage of the coupled-channel $T$-matrix  
is a significant  step toward realistic analyses of  $S$=$-2$ hypernuclear systems.

We are  also interested in the fact that J\"ulich Ch-EFT and HAL QCD give quite similar $\Xi N$ phase shifts throughout 
the $S$-wave spin- and isospin channels~\cite{JEF,HAL2}.  In particular, they both predict a structure  close to
the $\Xi N$ threshold that is related to the coupling to the $\Lambda\Lambda$ state.
Thus, we first describe   in detail the $S$=$-2$ interactions employed in  Sect.~\ref{two body}.
The $\Xi NN$ Faddeev equations and  the results are presented in Sect.~\ref{faddeev}.
%
%
%
\begin{figure}[ht]
\centering
  \includegraphics[clip, width=110mm]{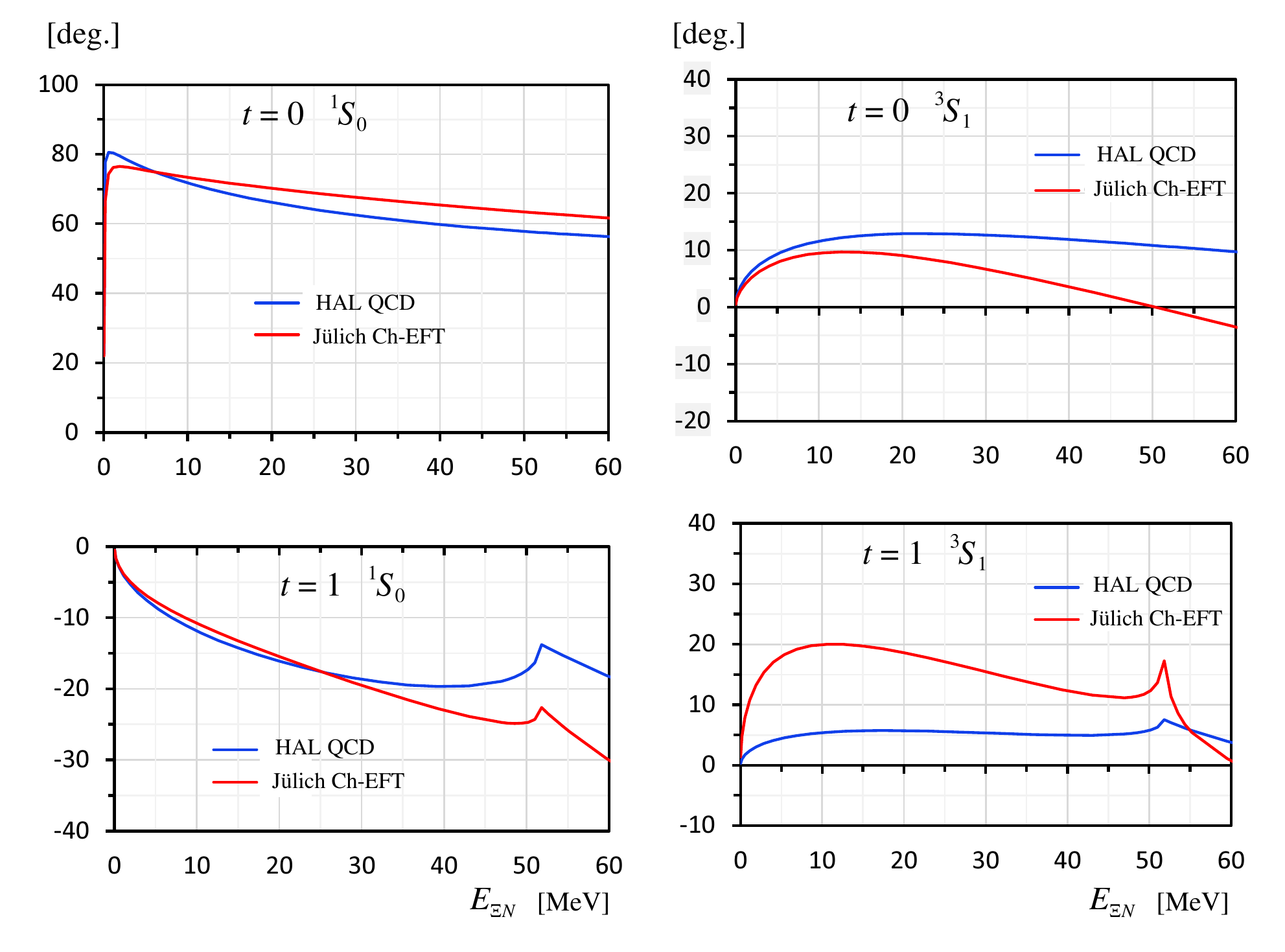}
  \caption{$\Xi N$ $S$-wave phase shifts generated by J\"ulich Ch-EFT (red lines)  and HAL QCD (blue lines) 
  for isospin $t=0$ and $t=1$ states.
                 }
  \label{fig1}
\end{figure}
%
\section{ $S$=$-2$ baryon-baryon interactions employed}
\label{two body}
\begin{figure}
\centering
  \includegraphics[clip, width=110mm]{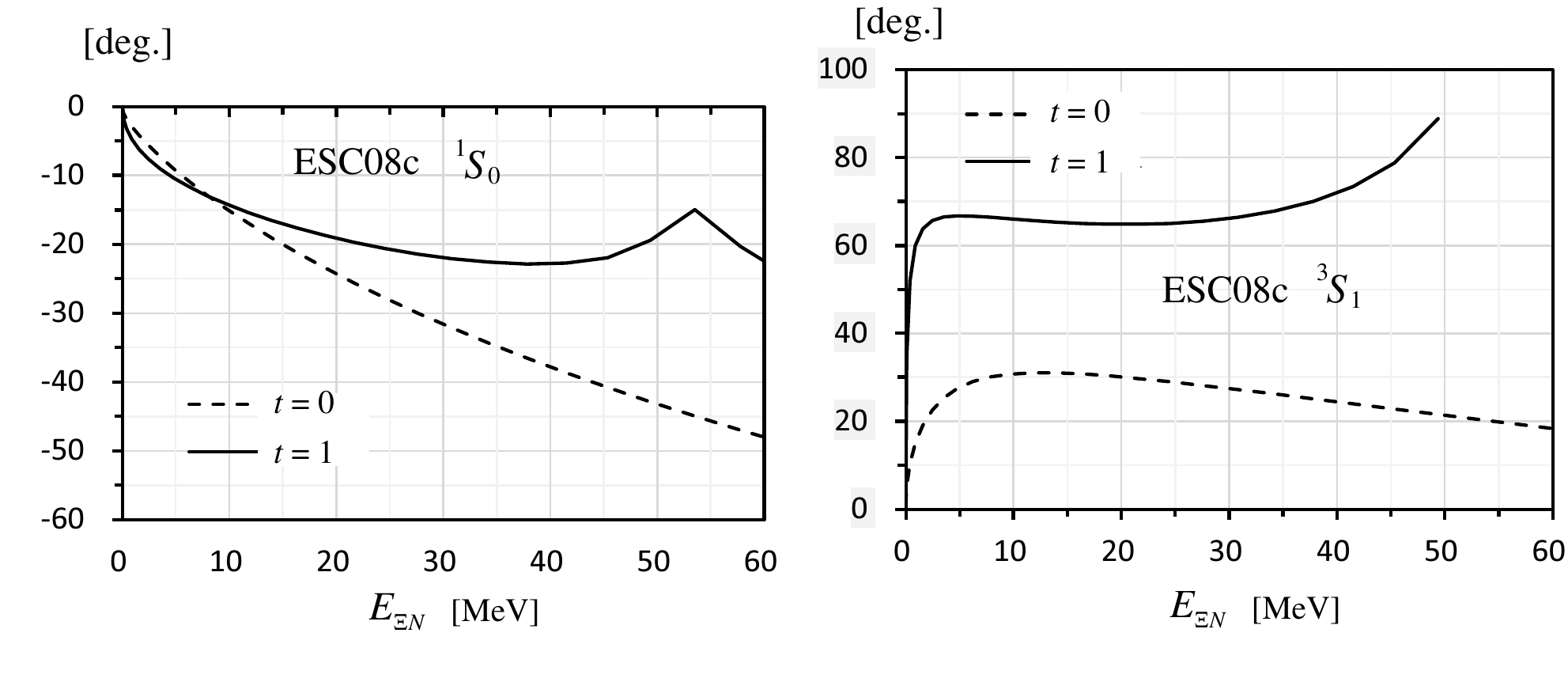}
  \caption{$\Xi N$ $^1S_0$ (left panel) and $^3S_1$ (right panel) phase shifts generated by ESC08c for isospin 
                 $t=0$ (dashed lines)   and $t=1$ (solid lines). }
  \label{fig2}
\end{figure}

In Fig.~\ref{fig1},
$\Xi N$ $S$-wave phase shifts generated by J\"ulich Ch-EFT and  HAL QCD are shown for isospin $t=$0 and $t=$1 states.
Interestingly, these two 
interactions give quite similar phase shifts except at lower energies of the $^3S_1$, $t=$1 state.
The $^1S_0$ phase shift for $t=$0  to which the $\Lambda\Lambda$ channel is coupled  also shows noticeable behavior; 
it indicates a strongly attractive feature quickly rising up to 80 degrees from the threshold.
In contrast to this, ESC08c gives repulsive phase shift for the $^1S_0$, $t=$0 state as shown in Fig.~\ref{fig2}. 
A characteristic of ESC08c is that a bound state exists in $^3S_1$-$^3D_1$ for $t =$1~\cite{ES8}. 
However, as realized from  Fig.~\ref{fig2},  the force used here for this partial wave is not so attractive as it generates
a bound state. 
The numerical code used~\cite{Yam} is given by one of the authors of Ref.~\cite{ES8}, and we examine rigorously 
the phase shifts shown in  Fig.~\ref{fig2}.  
Thus, ESC08c employed in this paper is not identical to the original, but nevertheless in Sec.~\ref{faddeev}, 
the three-body results for ESC08c will be shown for reference. 

\begin{figure}
\centering
  \includegraphics[clip, width=80mm]{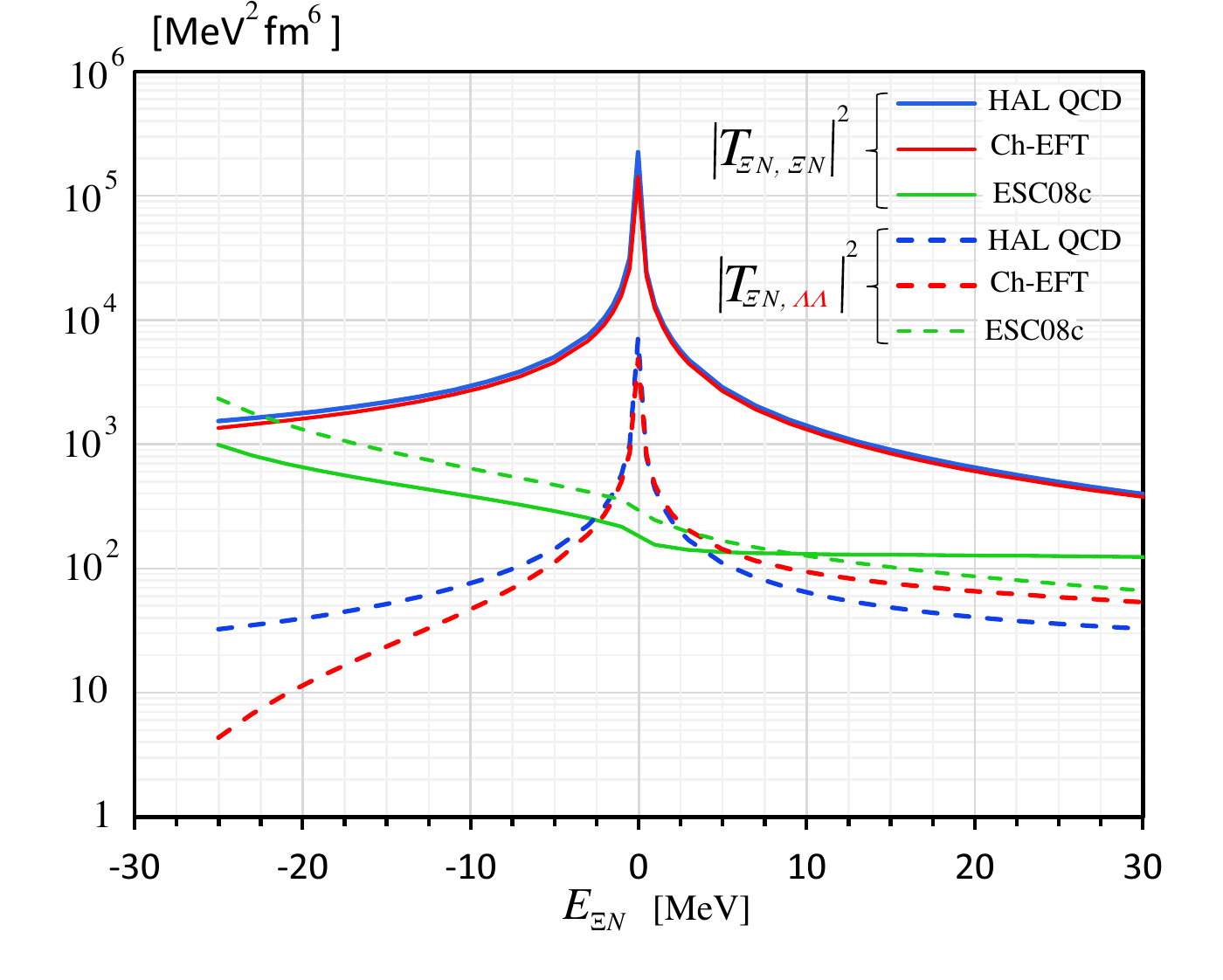}
  \caption{Absolute squares of $T_{\Xi N , \Xi N}$ and $T_{\Xi N, \Lambda\Lambda}$ for the $^1S_0$, $t=0$ channel generated 
               by J\"ulich Ch-EFT (red lines), HAL QCD(blue lines) and ESC08c(green lines).  }
  \label{fig3}
\end{figure}

\begin{figure}
\centering
  \includegraphics[clip, width=120mm]{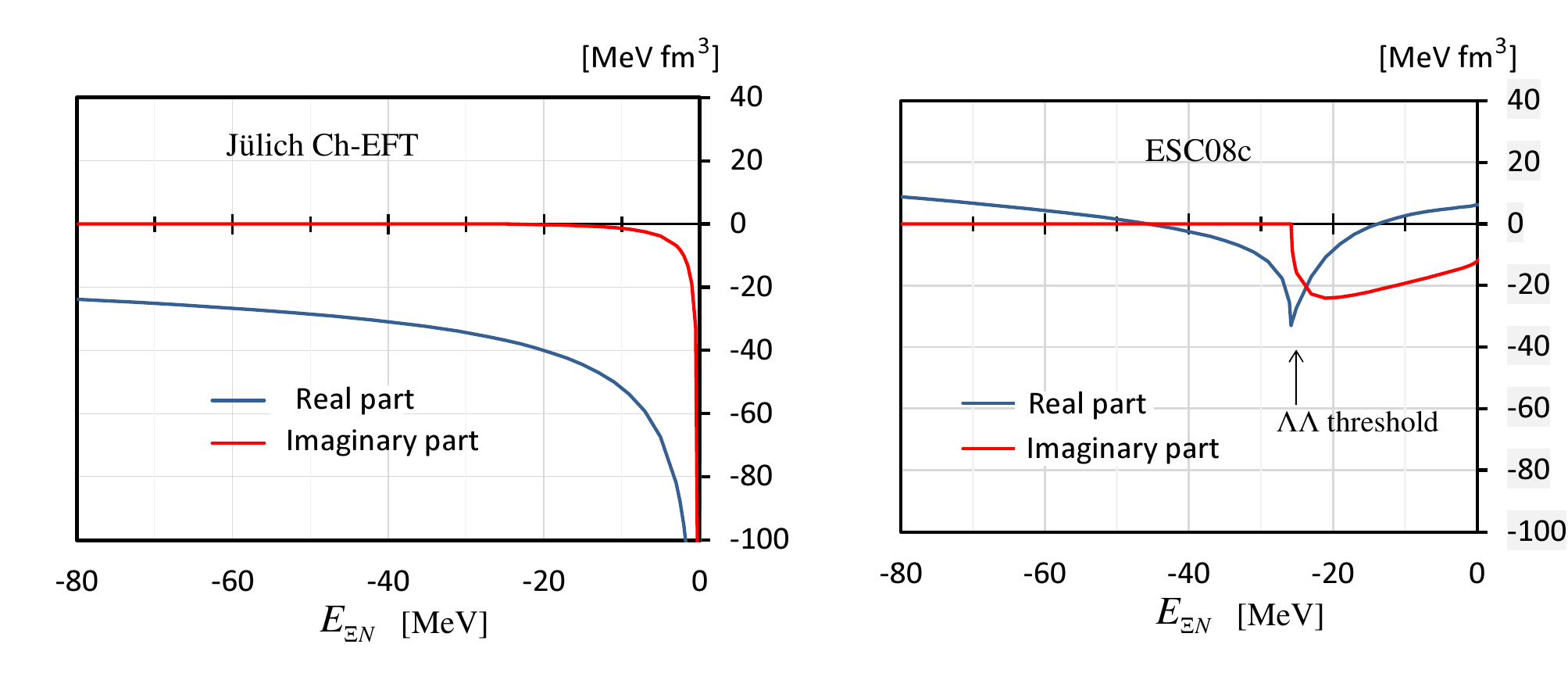}
  \caption{Real and imaginary parts of  $T_{\Xi N , \Xi N}$  for the $^1S_0$, $t$=0 channel  below the $\Xi N$ threshold
  depicted by blue and red lines respectively, which
are generated by J\"ulich Ch-EFT  (left panel) and ESC08c (right panel).
 }
  \label{fig4}
\end{figure}

Since the $\Xi N$ state can decay into $\Lambda\Lambda$ in $^1S_0$ for  $t=$0, we closely investigate the $T$-matrices 
for this channel. 
In Fig.~\ref{fig3},  both of $|T_{\Xi N , \Xi N}|^2$ and $|T_{\Lambda\Lambda , \Xi N}|^2$ by J\"ulich Ch-EFT and HAL QCD
show visible cusps just at the $\Xi N$ threshold, which are caused by an inelastic virtual-state pole 
close to the threshold~\cite{JEF,HAL2}.  In more detail, the real and the imaginary parts of  $T_{\Xi N , \Xi N}$ 
below the $\Xi N$ threshold are illustrated in Fig.~\ref{fig4}.
As indicated on the left panel,  the magnitude of the imaginary part of J\"ulich Ch-EFT is negligibly small compared to 
that of the real part. This is the reason why we utilize only the real part of  $T_{\Xi N , \Xi N}$ as an input to the $\Xi NN$
Faddeev calculation, and treat it as a bound state problem.  By contrast, the imaginary part of ESC08c has a significant 
magnitude indicated by  the red line on the right panel, which makes us unable to address the $\Xi NN$ system 
as a bound state. 
%
%
\section{ $\Xi NN$ Faddeev equation and Results}
\label{faddeev}
The Faddeev equations for the system consist of two nucleons and a hyperon can be seen in many literatures.
Let us now assign the number 1 to  $\Xi$, 2 and 3 to two nucleons, and impose antisymmetry to the total wave
function $\Psi$\ :  \ $P_{23}\, \Psi=-\Psi$ where $P_{23}$ is the transposition operator for two nucleons.  Then the Faddeev
components for the bound-state problem satisfy the coupled equations,
\begin{eqnarray}
\psi^{(23)} =& G_0 \ \, t_{NN}
           \  (1-P_{23}) \ \psi^{(12)}  \hskip10pt
\nonumber
\\[5pt]
\psi^{(12)} =& G_0 \ \,  t_{\Xi N} 
    \ (\psi^{(23)} -P_{23} \ \psi^{(12)} )
  \label{eq1}
\end{eqnarray}
where $\Psi=\psi^{(23)}+(1-P_{23}) \ \psi^{(12)}$.   To solve this coupled set, we follow the way used in~\cite{MaG,MaG2};  
 \, the set~(\ref{eq1}) of integral equations is put into the form 
\begin{equation}
\eta (E) \, \tilde\psi = \tilde K (E) \, \tilde \psi
  \label{eq2}
\end{equation}
with $\eta (E)$ added to the left, and this eigenvalue problem is solved at a fixed energy $E$ below the $\Xi d$ 
threshold.   If a bound state exists,  an eigenvalue such as $\eta (E_b)$=1  can be found at the bound-state energy
 $E_b$.

\begin{figure}[h]
\centering
  \includegraphics[clip, width=70mm]{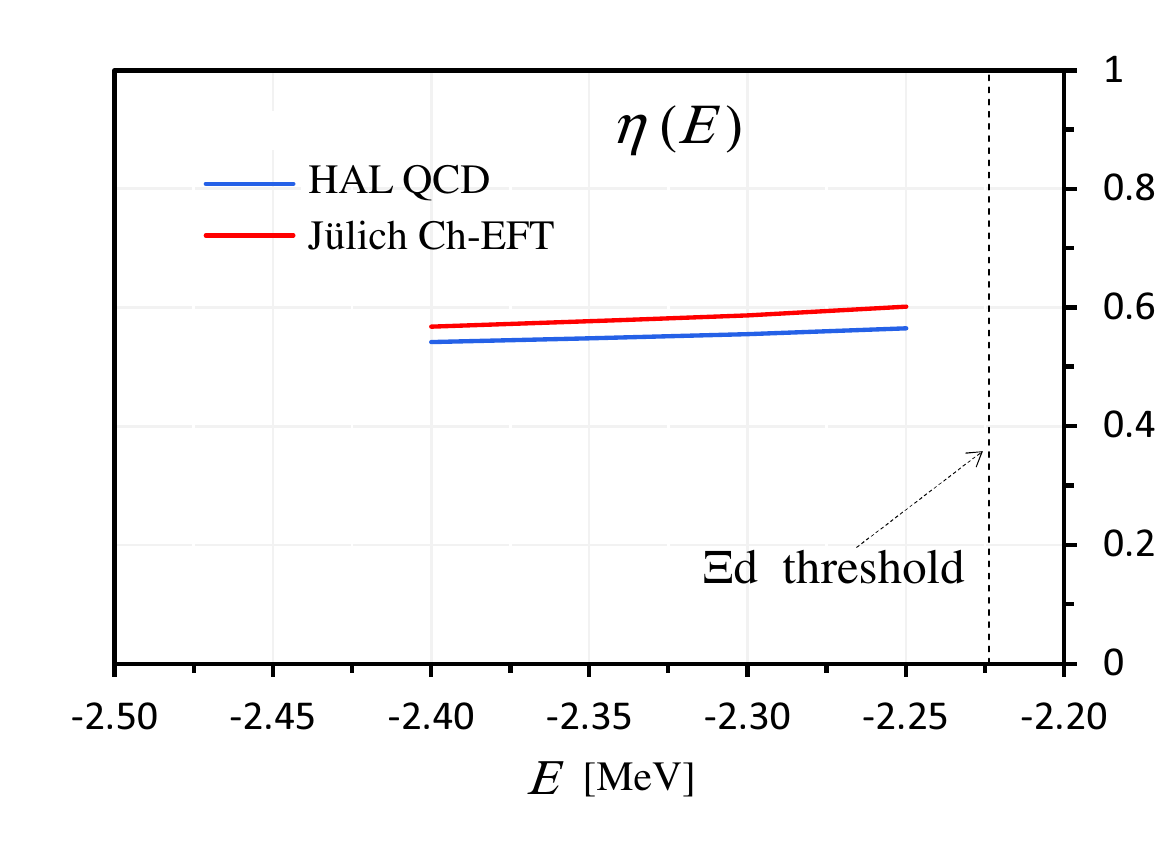}
  \caption{Eigenvalue $\eta (E)$ of the Faddeev kernel $\tilde K (E)$ as a function $E$ below the $\Xi d$ threshold.  
             The energy $E$ is set to zero at the $\Xi NN$ threshold.  
 The red and blue lines indicate the values generated by J\"ulich Ch-EFT  and HAL QCD respectively.}
  \label{fig5}
\end{figure}

We first analyze  the state with the total isospin and spin-parity  $(T,J^{\pi})=(1/2, 1/2^+)$,  which is most likely
bound owing to the contribution  from the  $\Xi $-deuteron  configuration. 
As mentioned in Sect.~\ref{two body},  the calculations are performed only for J\"ulich Ch-EFT  and HAL QCD,
where the imaginary part of  $T_{\Xi N, \Xi N}$ is negligibly small below the threshold and only the real part is incorporated.
In Fig.~\ref{fig5},  eigenvalue $\eta (E)$ is shown as a function of $E$ below the $\Xi d$ threshold at $-2.225$ MeV.
The energy $E$ is set to zero at the $\Xi NN$ threshold.
Although  the $\Xi N$ phase shifts in the  $t=$0, $^1S_0$  state show a strongly attractive feature,  the eigenvalues are
far from $\eta (E)=$1, thus the $\Xi NN$ system is not bound.  Through detailed investigations, we confirm that 
the $\Xi N$,  $t=$1 force  for $^1S_0$  which shows repulsive behavior in Fig.~\ref{fig1}  prevents the state from binding.
%
%

We also study the $(T,J^\pi)=(1/2, 3/2^+)$ state.  In this case,  the overlap
of  the  $\Xi N$,  $^1S_0$  state with the total spin $J^\pi= 3/2^+$ in angular-momentum coupling is negligible, and decays
into $\Lambda\Lambda N$ are suppressed. Hence, we  perform bound-state calculations for ESC08c 
in addition to J\"ulich Ch-EFT  and HAL QCD.  
No bound state is found  also in this   $(T,J^\pi)=(1/2, 3/2^+)$ state for J\"ulich  Ch-EFT and HAL QCD,  but a bound state
exists at $E=-3.05$ MeV for ESC08c.  The attraction in the $\Xi N$, $^3S_1$ state for  t=1  shown in Fig.~\ref{fig2} is the main
contribution
to this binding.

In conclusion, we have performed  the $\Xi NN$ bound-state calculations for  the  $(T,J^\pi)=(1/2, 1/2^+)$ and 
$(T,J^\pi)=(1/2, 3/2^+)$ states using the  coupled-channel $T$-matrix $T_{\Xi N, \Xi N}$ with negligible imaginary
parts below the threshold for J\"ulich Ch-EFT and HAL QCD.  It turns out that no bound state exists.  In spite of
the  $\Xi N$ strong attraction in the  $^1S_0$, $t=$0 state,  repulsive effects from the isosin partner,  $^1S_0$,  $t=$1
state prevent the binding.   In contrast,  ESC08c generates a bound state at
$E=-3.05$ MeV  for  $(T,J^\pi)=(1/2, 3/2^+)$ where the decays into $\Lambda\Lambda N$ are suppressed owing
to negligible angular-momentum coupling.  This is brought about
by an attractive feature of  the $^3S_1$,  $t=$1 state for ESC08c.

\begin{acknowledgements}

We thank Y. Yamamoto,  J. Haidenbauer  and T. Inoue  for  the communication with regard to
their $S=-2$ baryon-baryon interactions.

\end{acknowledgements}



\end{document}